\documentclass
[preprint, showpacs,preprintnumbers,amsmath,amssymb,nofootinbib]{revtex4}
\usepackage{amssymb}
\usepackage{natbib}
\usepackage{graphicx}
\usepackage{dcolumn}
\usepackage{bm}

\newcommand{\bea}{\begin{eqnarray}}
\newcommand{\eea}{\end{eqnarray}}

\begin{document}
\title{ Testing the cosmic distance duality relation using Type Ia supernovae and BAO observations }
\author{Fan Yang$^1$, Xiangyun Fu$^1$\footnote{corresponding author: xyfu@hnust.edu.cn}, Bing Xu$^2$\footnote{corresponding author: xub@ahstu.edu.cn},
Kaituo Zhang$^3$, Yang Huang$^1$,
and Ying Yang$^1$}
\address{$^1$Department of Physics, Key Laboratory of Intelligent Sensors and Advanced Sensor Materials, Hunan University of Science and Technology, Xiangtan, Hunan 411201, China\\
$^2$School of Electrical and Electronic Engineering, Anhui Science and Technology University, Bengbu, Anhui 233030, China\\
$^3$Department of Physics, Anhui Normal University, Wuhu, Anhui 241000, China}

\begin{abstract}
In this work, we propose to utilize the observed ratio of spherically-averaged distance to the sound horizon scale from Baryon Acoustic Oscillation (BAO) data to test the cosmic distance duality relation (CDDR) by comparing the luminosity distances (LDs) obtained from Type Ia supernovae (SNIa) observations with angular diameter distances (ADDs) derived from these ratio measurements, using a cosmological-model-independent method. To match the LDs with the ADDs at the identical redshifts, we employ two methods: a compressed form of the Pantheon sample and a hybrid approach that combines the binning method with an artificial neural network (ANN). The Hubble parameter $H(z)$ at any redshift is reconstructed from the observed Hubble parameter data with the ANN to derive the ADD. To avoid potential biases resulted from the specific prior values of the absolute magnitude $M_{\rm B}$ of SNIa and the sound horizon scale $r_{\rm d}$ from BAO measurements, we introduce the fiducial parameter $\kappa\equiv10^{M_{\rm B} \over 5}\, r_{\rm d}^{3 \over 2} $ and marginalize their impacts by treating them as nuisance parameters with flat prior distributions in our statistical analysis. Subsequently, we update the measurements of ratio of the transverse comoving distance to the sound horizon scale from the latest BAO data released by the Dark Energy Spectroscopic Instrument (DESI) collaboration for CDDR testing. Our results indicate that BAO observation provides a powerful tool for testing the CDDR, independent of both the absolute magnitude $M_{\rm B}$ and sound horizon scale $r_{\rm d}$, as well as any cosmological model.

$\mathbf{Keywords:}$ Cosmic distance duality relation, BAO observation, Cosmological-model-independent method
\end{abstract}


 \maketitle

\section{Introduction}
The cosmic distance duality relation (CDDR) is a fundamental relationship in modern cosmology. Specifically, it holds that at a given redshift, the relationship is expressed as ${D_{\rm L}}={D_{\rm A}}{(1+z)}^{2}$~\cite{Etherington1933,ellis2007}. The validity of the CDDR relies on three fundamental assumptions: the space-time is described by a metric theory, the light follows null geodesics between the source and observer, and that the number of photons is conserved. The CDDR finds applications across various astronomical domains, including the large-scale distribution of galaxies, the uniformity of the Cosmic Microwave Background (CMB) temperature~\cite{Aghanim2020}, the gas mass density and temperature profile of galaxy clusters~\cite{Cao2011,Cao2016}. However, a potential violation of any of the underlying assumptions of the CDDR could indicate the presence of exotic physics~\cite{Bassett2004a,Bassett2004b}. Hence, it is imperative to test the CDDR with various observational data and reliable methods.

A considerable amount of research has been conducted to test the CDDR using various astronomic observations~\cite{Holanda2010,Holanda2019,Holanda2022,Li2011,Liang2013,Lima2021,Wu2015,Ruan2018,Liao2022,Li2018,Xu2022,Liao2016,Tang2023,Liu2023a,Lin2020,Lin2021,Arjona2021}. Type Ia supernovae (SNIa) are widely utilized for determining luminosity distance (LD) $D_{\rm L}(z)$. Meanwhile, angular diameter distance (ADD) $D_{\rm A}(z)$ is frequently derived through multiple observational techniques, encompassing the Sunyaev-Zeldovich effect and gas mass fraction measurements in galaxy clusters~\cite{Holanda2010,Li2011,Liang2013,Lima2021}, baryon acoustic oscillations (BAO)~\cite{Wu2015}, strong gravitational lensing (SGL) systems~\cite{Ruan2018,Liao2022}, and the angular size of ultra-compact radio sources~\cite{Li2018}. The findings indicate that the CDDR aligns with current astrophysical observations across various redshift ranges, as reported in Refs.~\cite{Avgoustidis2010,Stern2010,Holanda2012a,Holanda2013,Liao2011,Holanda2017a,Fu2011,Fu2017,Zheng2020}.

Due to the lack of astronomical observational data, it is currently difficult to obtain the LD and ADD at the same redshift from a single astronomical observation. To acquire LD and ADD from astrophysical observations at the same redshifts, researchers such as Holanda {\it et al.}~\cite{Holanda2010} and Li {\it et al.}~\cite{Li2011} have utilized galaxy cluster samples~\cite{Bonamenteet2006,DeFilippis2005} and SNIa data to find the closest measurements (with a redshift difference of less than 0.005) for CDDR validation. To reduce potential statistical inaccuracies arising from the use of a single SNIa data point among those meeting the selection criteria, Meng {\it et al.}~\cite{Meng2012} employed a binning method, aggregating the eligible data into bins to calculate LD.

SNIa and BAO observations play important roles in testing the CDDR. Recently, Wu {\it et al.} examined the CDDR through a comparison of the Union2.1 dataset with the five ADD values derived from BAO observations, concluding that the high precision of BAO measurements makes them an effective tool for validating the CDDR~\cite{Wu2015}. Notably, the LD derived from SNIa observations depends on the peak absolute magnitude $M_{\rm B}$ of these supernovae, which is traditionally considered a fixed value, independent of any other factors. However, recent research has focused on deducing $M_{\rm B}$ from a cosmological perspective~\cite{Camarena2020a,Dinda2023}. Variations in $M_{\rm B}$ are derived from SNIa observation like Pantheon, combined with other data sets, including cosmic microwave background (CMB) observations, cosmic chronometer Hubble parameter data, and BAO observations. Discrepancies in the Cepheid-calibrated absolute magnitude of SNIa are observed between redshifts $z\leq 0.01$ and $z>0.01$~\cite{Camarena2021,Camarena2020b}. For example, the CMB constraints on the sound horizon predicts $M_{\rm B}$ to be approximately $M_{\rm B}\sim -19.4\,{\rm mag}$ using an inverse distance ladder~\cite{Dinda2023}, while the SH0ES estimates $M_{\rm B}$ to be around $M_{\rm B}\sim -19.2\,{\rm mag}$~\cite{Camarena2020a}. Recent studies Refs.~\cite{Kazantzidis2021,Kazantzidis2020} have also suggested the possibility of a weak evolution in $M_{\rm B}$.

In addition, the fitting issue remains to pose a challenge in utilizing the BAO peak position as a cosmological standard ruler, despite the fact that BAO measurements are instrumental in probing a multitude of cosmological parameters. Specifically, Roukema {\it et al.} have recently identified a dependence of the BAO peak location on the surrounding environment~\cite{Roukema2015,Roukema2016}. Ding {\it et al.} and Zheng {\it et al.} identified a significant systematic discrepancy between Hubble $H(z)$ measurements derived from BAO and those from differential aging (DA) techniques~\cite{Ding2015,Zheng2016}. The distinct sound horizon scales $r_{\rm d}$ and the current value of the Hubble constant $H_0$ are derived from a diverse array of observational data sources. These include CMB observations~\cite{Ade2016,Bennett2013}, the Sloan Digital Sky Survey (SDSS) data release 11 galaxies~\cite{Carvalho2020}, and BAO measurements~\cite{Verde2017}. For instance, utilizing the SDSS data release 11 galaxies and priors on the matter density parameter from SNIa data, Carvalho {\it et al.} derived constraints on $r_{\rm d}$ at relatively low redshifts~\cite{Carvalho2020}, with $r_{\rm d}={107.4\pm1.7\,h^{-1}{\rm Mpc}}$. Here, $h$ represents the Hubble constant $H_0$ in units of $100 {\rm km\,s^{-1}Mpc^{-1}}$. Additionally, Verde {\it et al.} also measured the sound horizon using SNIa and BAO data~\cite{Verde2017}, yielding $r_{\rm d}={101.0\pm2.3{\, h^{-1}}{\rm Mpc}}$. As a result, CDDR tests that depend on the priors of $M_{\rm B}$ and $r_{\rm d}$ are not completely independent of cosmological model assumptions. This is because the method used to derive the LD and ADD, which is based on the value of $M_{\rm B}$ of SNIa observations and $r_{\rm d}$ of BAO measurements, displays a certain level of dependency on the cosmological model. Ma {\it et al.} utilized a Bayesian approach to estimate the SNIa luminosity distance moduli at redshifts corresponding to ADD data from BAO observations~\cite{Ma2018}. Although the mentioned method enhances testing precision, it still requires the assumption of a cosmological model to obtain values for $M_{\rm B}$ and $r_{\rm d}$. Moreover, such priors could introduce biases into CDDR tests. More recently, Jesus {\it et al.} conducted a cosmographic analysis utilizing the Pad$\acute{\rm e}$ method, incorporating data from the Pantheon Plus SNIa samples, baryon acoustic oscillations from SDSS and DESI, as well as cosmic chronometers, to test the CDDR~\cite{Jesus2024}. Their findings revealed no violation of the CDDR. Nevertheless, the results suggested a preference for varying $M_{\rm B}$ and $H_0$ values when different combinations of SNIa, Hubble, and BAO observations were considered, indicating that potential validations of the CDDR introduce new statistical correlations among cosmographic parameters.   Hence, it is significant to seek new methods for CDDR validation that are independent of these parameters.

Recently, Xu {\it et al.} introduced an approach in Ref.~\cite{Xu2022} to test the CDDR by marginalizing over $M_{\rm B}$ and $r_{\rm d}$. This was done by comparing the five BAO measurements of $D_{\rm M}(z)/r_{\rm d}$ from the extended Baryon Oscillation Spectroscopic Survey (eBOSS) DR16 quasar dataset with the Pantheon SNIa sample. Here, $D_{\rm M}$ denotes the transverse comoving distance. This methodology eliminated the need for assuming a cosmological model and removed the dependence on the specific prior values of $M_{\rm B}$ and $r_{\rm d}$. Subsequently, Wang {\it et al.}~\cite{Wang2024} demonstrated that specific prior values of $M_{\rm B}$ and $r_{\rm d}$ lead to significant biases
on the CDDR test by using 13 transverse BAO measurements from the SDSS with the Pantheon SNIa samples. To circumvent these biases, they propose a method independent of $M_{\rm B}$ and $r_{\rm d}$ to test CDDR by considering the fiducial value of $\kappa\equiv10^{M_{\rm B} \over 5}r_{\rm d}$ as a nuisance parameter and then marginalizing its influence with a flat prior in the analysis.

It is acknowledged that the BAO observational data comprises three types of measurements: the ratios of transverse comoving distance $D_{\rm M}(z)$, the Hubble distance $D_{\rm H}(z)$, and the spherically-averaged distance $D_{\rm V}(z)$ to the sound horizon scale $r_{\rm d}$. Since the measurements of $D_{\rm V}(z)/r_{\rm d}$ encompass information regarding the transverse comoving distance $D_{\rm M}(z)$, BAO observations present an additional avenue for acquiring ADD, thereby enabling the testing of the CDDR. Furthermore, the Dark Energy Spectroscopic Instrument (DESI) collaboration has recently Released Data release 1 (DR1) with BAO measurements~\cite{Adame2024a,Adame2024b}. Presently, there are five BAO data points for the measurements of $D_{\rm V}(z)/r_{\rm d}$ and nine for the measurements of $D_{\rm M}(z)/r_{\rm d}$ BAO data points. To ensure and uphold the integrity and accuracy of BAO observational data, it is meaningful to test the CDDR by utilizing the latest DESI BAO data, specifically focusing on the measurements of $D_{\rm V}(z)/r_{\rm d}$. This constitutes the principal motivation for the present work.

In this research, we conduct a validation of the CDDR by comparing the LD inferred from the SNIa against the ADD derived from BAO. The function $\eta(z)=D_{\rm L}(z)(1 + z)^{-2}/D_{\rm A}(z)$ verifies potential deviations at all redshifts. To achieve alignment between the SNIa and BAO at identical redshifts, we employ the Artificial Neural Network (ANN) integrated with the binning method and a compressed version of the Pantheon sample. First, the ADD is inferred from the spherically-averaged distance $D_{\rm V}(z)$. To obtain ADD from $D_{\rm V}(z)/r_{\rm d}$ BAO observations, we reconstruct Hubble parameter data from the DA technique with an ANN. To avoid potential bias in the CDDR test arising from the particular prior values of the absolute magnitude $M_{\rm B}$ of SNIa and the sound horizon scale $r_{\rm d}$ from BAO measurements, we introduce a new variable $\kappa\equiv10^{M_{\rm B} \over 5}\, r_{\rm d}^{3 \over 2} $. The variable is treated as nuisance parameters with a flat prior distribution in our statistical analysis. Then, we updated the $D_{\rm M}(z)/r_{\rm d}$ dataset for testing the CDDR with the latest BAO observations from the DESI collaboration. Our findings demonstrate that the CDDR is consistent with observations and that our method for testing the CDDR is independent not only of specific cosmological models but also of the particular prior values of $M_{\rm B}$ and $r_{\rm d}$.

\section{Data and Methodology}

\subsection{Data}
In this study, we utilize the sample of SNIa from the Pantheon dataset, encompassing 1048 SNIa within the redshift range of $0.01 < z < 2.3$, as provided by the Pan-STARRS1 (PS1) Medium Deep Survey~\cite{Scolnic2018}. SNIa serve as the most immediate evidence for the accelerating expansion of the universe, acting as standard candles. The Pantheon dataset yields a distance modulus that has been calibrated via the SALT2 light-curve fitting procedure, employing the Bayesian Estimation Applied to Multiple Species with Bias Corrections (BEAMS) method to ascertain the nuisance parameters and to correct for distance biases. This calibration entails the utilization of the distance modulus formula, $\mu = m_{\rm B}-M_{\rm B}$, in which $m_{\rm B}$ denotes the observed peak apparent magnitude within the B-band of the rest frame. The correlation between the LD $D_{\rm L}$~\cite{Zhou2019} and the $\mu$ can be expressed as follows
\begin{equation}
\mu(z)=5\log_{10}(D_{\rm L}(z))+25\,.
\label{muz}
\end{equation}

\begin{table}[t]
\begin{tabular}{| c | ccc c |}
         \hline
	Measurement & $z_\mathrm{BAO}$ & Value & Method II & Ref. \\ \hline
    6dFGS & 0.106 & $D_{\rm V}(z)/r_{\rm d}=3.06\pm0.14$ & Binning & \cite{Beutler2011}\\
	SDSS DR7 MGS & 0.15 & $D_{\rm V}(z)/r_{\rm d}=4.47\pm0.16$ & Binning & \cite{Ross2015} \\
	SDSS DR14 eBOSS Quasar & 1.52 & $D_{\rm V}(z)/r_{\rm d}=26.0\pm 1.0$ & ANN & \cite{Ata2018}\\
    DESI DR1 BGS & 0.30 & $D_{\rm V}(z)/r_{\rm d}=7.93\pm0.15$ & Binning & \cite{Adame2024c} \\
    DESI DR1 Quasar & 1.49 & $D_{\rm V}(z)/r_{\rm d}=26.07\pm0.67$ & ANN & \cite{Adame2024c} \\

	\hline
    SDSS DR12 BOSS Galaxy & 0.38 & $D_M/r_d=10.23\pm0.17$ & Binning & \cite{Alam2017}\\
    SDSS DR12 BOSS Galaxy & 0.51 & $D_M/r_d=13.36\pm0.21$ & Binning & \cite{Alam2017}\\
    SDSS DR16 eBOSS LRG & 0.70 & $D_M/r_d=17.86\pm0.33$ & Binning & \cite{Gil2020}\\
    SDSS DR16 eBOSS ELG & 0.85 & $D_M/r_d=19.5\pm1.0$ & Binning & \cite{Tamone2020} \\
    SDSS DR16 eBOSS Quasar & 1.48 & $D_M/r_d=30.69\pm0.80$ & ANN & \cite{Neveux2020}\\
    DESI DR1 LRG & 0.51 & $D_M/r_d=13.62\pm0.25$ & Binning & \cite{Adame2024c} \\
    DESI DR1 LRG & 0.71 & $D_M/r_d=16.85\pm0.32$ & ANN & \cite{Adame2024c} \\
    DESI DR1 LRG+ELG & 0.93 & $D_M/r_d=21.71\pm0.28$ & Binning & \cite{Adame2024c} \\
    DESI DR1 ELG & 1.32 & $D_M/r_d=27.79\pm0.69$ & ANN & \cite{Adame2024c} \\
	\hline
\end{tabular}
	\caption{List of BAO measurements used in this work, indicating the use of either the Binning method or ANN in method II.\label{tab:bao_measures}}
\end{table}

BAO denotes a pattern of excess density or aggregation of baryonic matter at specific length scales, resulting from the oscillations of acoustic waves that propagated through the early universe. These oscillations manifest on characteristic scales and offer a cosmological standard ruler for measuring length scales, thereby facilitating the investigation of the history about universe expansion. The extent of this standard ruler ($\sim$150 Mpc in the present universe) corresponds to the distance that sound waves emanating from a point source at the conclusion of inflation would have traversed prior to decoupling.

The measurements of $D_{\rm V}(z)/r_{\rm d}$ from BAO have been released from the 6-degree Field Galaxy Survey (6dFGS) at redshift ($z = 0.106$)~\cite{Beutler2011}, main galaxy sample (MGS) at redshift ($z = 0.15$)~\cite{Ross2015}, and quasars tracers at redshift ($z = 1.52$) from the SDSS~\cite{Ata2018}. Recently, the DESI collaboration~\cite{Aghamousa2016,Abareshi2022,Albrecht2006} published its DR1, which includes the two additional $D_{\rm V}(z)/r_{\rm d}$ measurements derived from the bright galaxy survey (BGS) and quasars as direct tracers, along with five measurements of $D_{\rm M}(z)/r_{\rm d}$ from emission line galaxies (ELG), Luminous Red Galaxies (LRG), and Lyman-$\alpha$ (Ly $\alpha$) forest quasar tracers~\cite{Adame2024a,Adame2024b}, as detailed in Tab. I of Ref.~\cite{Adame2024c}. As a result, we now have a combined set of five $D_{\rm V}(z)/r_{\rm d}$ measurements to derive the ADD, and use them to impose constraints on the CDDR. Additionally, Xu {\it et al.} have tested the CDDR~\cite{Xu2022} by utilizing the five $D_{\rm M}(z)/r_{\rm d}$ data points from the galaxy, LRG, ELG, and quasars tracers~\cite{Alam2021,Alam2017,Gil2020,Tamone2020,Neveux2020}, released by the eBOSS Collaboration. Furthermore, we update $D_{\rm M}(z)/r_{\rm d}$ measurements to test the CDDR by using a total of 9 $D_{\rm M}(z)/r_{\rm d}$ data points to calculate ADD. For a detailed overview of the data, please refer to Table~\ref{tab:bao_measures}.

The expressions for $D_{\rm M}(z)/r_{\rm d}$, $D_{\rm H}(z)/r_{\rm d}$ and $D_{\rm V}(z)/r_{\rm d}$ are written as:
\begin{equation}
{D_{\rm M}(z)/r_{\rm d}}={(1+z)D_{\rm A}(z)\over{r_{\rm d}}},
\end{equation}
\begin{equation}
{D_{\rm H}(z)/r_{\rm d}}={c\over {H(z)r_{\rm d}}},
\end{equation}
\begin{equation}
{D_{\rm V}(z)/r_{\rm d}}={[{zD_{\rm H}(z)D_{\rm M}(z)^2}]^{1/3}\over r_{\rm d}}\,.
\end{equation}
In the equation, $c$ is the speed of light, and $H(z)$ is the Hubble parameter. Therefore, the expression for $D_{\rm A}(z)$ given the observed data for $D_{\rm V}(z)/r_{\rm d}$ and $D_{\rm M}(z)/r_{\rm d}$, can be derived from the above equations, as follows:
\begin{equation}
{D_{\rm A}(z)={{{(D_{\rm V}(z)/r_{\rm d}})_{\rm obs}^{3/2}r^{3/2}_{\rm d}H(z)^{1/2}}\over c^{1/2}z^{1/2}(1+z)}},
\label{DA1}
\end{equation}
\begin{equation}
{D_{\rm A}(z)={{{(D_{\rm M}(z)/r_{\rm d})_{\rm obs}}r_{\rm d}}\over {1+z}}},
\label{DA2}
\end{equation}
Here, $(D_{\rm V}(z)/r_{\rm d})_{\rm obs}$ and $(D_{\rm M}(z)/r_{\rm d})_{\rm obs}$ denote the observed value of the ratio of spherically-averaged distance $D_{\rm V}$ and the transverse comoving distance $D_{\rm M}$, respectively, to the sound
horizon at the baryon drag epoch $r_{\rm d}$. It can be seen from the equation that the calculation of ADD $D_{\rm A}(z)$ using the $D_{\rm V}(z)/r_{\rm d}$ measurements necessitates the value of $H(z)$. In this work, the compilation of the Hubble parameter is obtained by utilizing the DA technique introduced in reference~\cite{Jimenez2002}. This method involves a comparative analysis of the ages of early-type galaxies that share similar metallicities and are distinguished by minute differences in redshift. This methodology effectuates a direct measurement of the $H(z)$ by employing the spectroscopic dating of passively evolving galaxies~\cite{Stern2010,Zhang2014,Moresco2012,Moresco2016,Ratsimbazafy2017,Moresco2015,Yang2023,
Marra2018,Wu2007}, ensuring its independence from any cosmological model. The compilation of 31 data points are also presented in Tab.~\ref{Hub31} for convenience.

To test the CDDR, it is necessary to match the LD values from the SNIa observation with the ADD values from BAO measurements at the same redshift. To achieve this, we employ two distinct methodologies to infer the apparent magnitude values at the redshifts of the BAO measurements.

\begin{table*}[htb]
\label{tab:results}
\begin{center}
\begin{tabular}{ccccc|cccc|ccccc} \hline \hline
& $z$ & $H(z)$ & $\sigma_{ H(z)}$ &  Ref. & $z$ & $H(z)$ & $\sigma_{ H(z)}$ &  Ref.  & $z$ & $H(z)$ & $\sigma_{ H(z)}$ &  Ref. \\  [1ex] \hline
& 0.07 & 69 & 19.6 & ~\cite{Zhang2014} & 0.4 & 95 & 17 & ~\cite{Stern2010} & 0.88 & 90 & 40 & ~\cite{Stern2010} \\ [1ex]
& 0.10  & 69 & 12 & ~\cite{Stern2010}  & 0.4004 & 77 & 10.2 & ~\cite{Moresco2016}  & 0.9  & 117 & 23 & ~\cite{Stern2010}     \\ [1ex]
& 0.12 & 68.6 & 26.2 & ~\cite{Zhang2014}  & 0.4247  & 87.1 & 11.2 & ~\cite{Moresco2016}  & 1.037 & 154 & 20 & ~\cite{Moresco2012}     \\ [1ex]
& 0.17 & 83 &  8 & ~\cite{Stern2010} & 0.4497 & 92.8 & 12.9 & ~\cite{Moresco2016} & 1.3 & 168 &  17 & ~\cite{Stern2010}    \\ [1ex]
& 0.1791 & 75 &  4 & ~\cite{Moresco2012} & 0.47 & 89 & 34 & ~\cite{Ratsimbazafy2017} & 1.363 & 160 &  33.6 & ~\cite{Moresco2015}    \\ [1ex]
& 0.1993 & 75 &  5 & ~\cite{Moresco2012} & 0.4783 & 80.9 & 9 & ~\cite{Moresco2016} & 1.43 & 177 &  18 & ~\cite{Stern2010}    \\ [1ex]
& 0.20 & 72.9 &  29.6 & ~\cite{Zhang2014} & 0.48 & 97 & 62 & ~\cite{Stern2010} & 1.53 & 140 &  14 & ~\cite{Stern2010}    \\ [1ex]
& 0.27 & 77 &  14 & ~\cite{Stern2010} & 0.5929 & 104 & 13 & ~\cite{Moresco2012} & 1.75 & 202 &  40 & ~\cite{Stern2010}    \\ [1ex]
& 0.28 & 88.8 &  36.6 & ~\cite{Zhang2014} & 0.6797 & 92 & 8 & ~\cite{Moresco2012} & 1.965 & 168.5 &  50.4 & ~\cite{Moresco2015}    \\ [1ex]
& 0.3519 & 83 &  14 & ~\cite{Moresco2012} & 0.7812 & 105 & 12 & ~\cite{Moresco2012}    \\ [1ex]
& 0.3802 & 83 & 13.5 & ~\cite{Moresco2016} & 0.8754 & 125 &  17 & ~\cite{Moresco2012}   \\ [-0.25ex]
\hline\hline
\end{tabular}
\caption[]{31 Hubble parameter measurements $H(z)$ obtained from the DA method (in units of ${\rm km \,s^{-1}Mpc^{-1}}$). }
\label{Hub31}
\end{center}
\end{table*}

\subsection{Compressed form of Pantheon sample}
We use a compressed form of the observational data from the Pantheon SNIa dataset, employing the technique suggested by Betoule {\it et al.}~\cite{Betoule2014}. In this approach, rather than aggregating the distance modulus as in Betoule {\it et al.}~\cite{Betoule2014}, the adjusted apparent magnitude is estimated by a piecewise linear function of $\mathrm{ln}(z)$, defined within each interval $z_{\rm b} \leq z \leq z_{\rm {b+1}}$ as follows:
\begin{equation}\label{IF}
\overline{m}_{\rm B}(z)=(1-\alpha)m_{{\rm B},{b}}+\alpha m_{{\rm B},{b+1}}\,.
\end{equation}
Within this framework, $\alpha$ is calculated as the logarithmic ratio of the redshift values, specifically $\alpha= \mathrm{ln}(z/z_{b} )/\mathrm{ln}(z_{{b+1}}/z_{b} )$, where $m_{{\rm{B}},{b}}$ denotes the apparent magnitude at the baseline redshift $z_{ b}$. To encompass the redshift limits of Pantheon sample, a set of 36 logarithmically spaced control points $z_{b}$ is established across the interval $0.01<z<2.3$. Subsequently, an interpolation methodology is engaged to fit the Pantheon dataset, culminating in a compressed form through the minimization of the $\chi^2$ function,
\begin{equation}
\chi^2 = \left[\mathbf{m}^*_{\rm B}-\mathbf{\overline{m}}_{\rm B}\right]^{\rm T} \cdot  \mathbf{Cov}^{-1} \cdot \left[\mathbf{m}^*_{\rm B}-\mathbf{\overline{m}}_{\rm B}\right].
\end{equation}

In this context, the covariance matrix is written as $\mathbf{Cov}=\mathbf{D}_{\rm stat}+\mathbf{C}_{\rm sys}$, where $\mathbf{D}_{\rm stat}$ denotes the statistical matrix characterized by owning only diagonal elements and $\mathbf{C}_{\rm sys}$ refers to the systematic covariance. Further details regarding the compressed form can be found in the Ref.~\cite{Xu2022}. In contrast to the conventional method of interpolating based solely on the closest pair of observations, the analysis derives the values of $\overline{m}_{\rm B}$ by incorporating the cumulative effect of all observations within the two adjacent redshift brackets of the target redshift. This approach substantially solves the detrimental effects that typically arise from a reduced dataset, notably the significant uncertainty. We refer to this approach using the Compressed form of the Pantheon sample as method I.

\subsection{Binning method and Artificial Neural Network}
To assess the validity of the CDDR, a direct approach is comparing ADD and LD sourced from distinct observational datasets at identical redshifts. Given the scarcity of ADD and LD observational data at the same redshift, we bin the LDs from SNIa data points that meet the selection criterion $\Delta z=|z_{\rm ADD}-z_{\rm SNIa}|<0.005$, following the methodology outlined in Refs.~\cite{Holanda2010,Li2011,Liao2016}. This technique, referred to as the binning method, reduces statistical errors that may arise from relying on a single SNIa data point within the selected dataset and has been previously utilized in the discourse on CDDR test as referenced in Refs.~\cite{Meng2012,Wu2015}. In this analysis, we compute the inverse variance-weighted mean of all the data points that meet the selection criteria. The determination of the weighted mean $\bar{m_{\rm B}}$ and its associated error $\sigma_{\bar{m_{\rm B}}}$ can be achieved through the application of standard data analysis methodologies as outlined in Chapter 4 of Ref.~\cite{Bevington1993},
\begin{equation}
\label{avdi1}
\bar{m_{\rm B}}={\sum(m_{{\rm B}i}/\sigma_{m_{{\rm B}i}}^2)\over \sum1/\sigma_{m_{{\rm B}i}}^2},
\end{equation}
\begin{equation}
\label{erroravdi1}
\sigma^2_{\bar{m_{\rm B}}}={1\over \sum1/\sigma_{m_{{\rm B}i}}^2}\,.
\end{equation}
Here, ${m_{{\rm B}i}}$ represents the $i$th data point of apparent magnitude, while $\sigma_{m_{{\rm B}i}}$ is associated with the respective observational error. Only three BAO data points with $D_{\rm V}(z)/r_{\rm d}$ measurements and six data points with $D_{\rm M}(z)/r_{\rm d}$ measurements satisfy the selection criteria. In order to use all BAO data to test the CDDR, we utilize the ANN to reconstruct a continuous $m_{{\rm B}}(z)$ function based on observations from the Pantheon SNIa dataset. Thus, for the BAO data that do not meet the selection criteria of the binning method but fall within the redshift range of the Pantheon SNIa survey, the ADD derived from these BAO data can be matched with the LD of SNIa at the same redshift.

\begin{figure}[htbp]
\includegraphics[width=8cm]{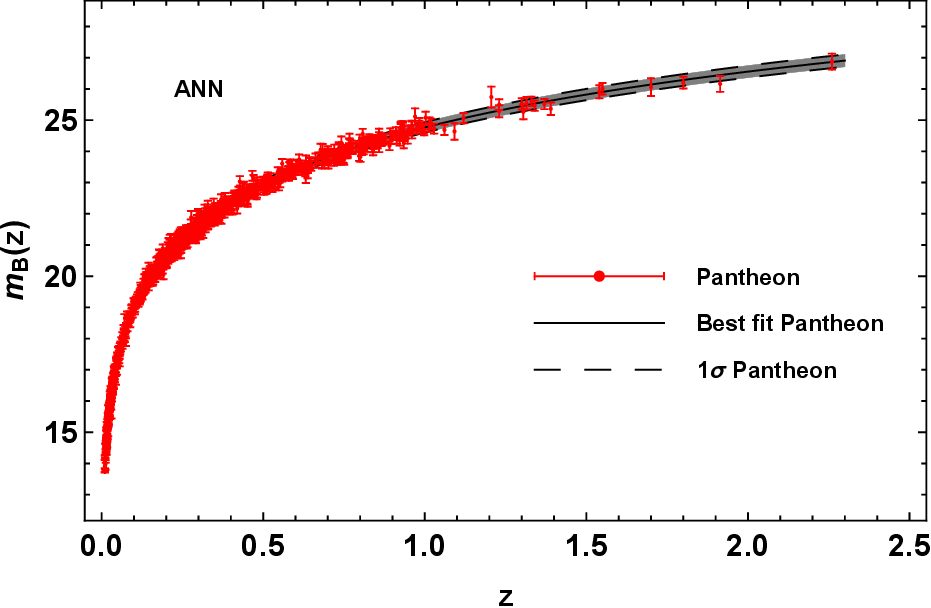}
\includegraphics[width=8cm]{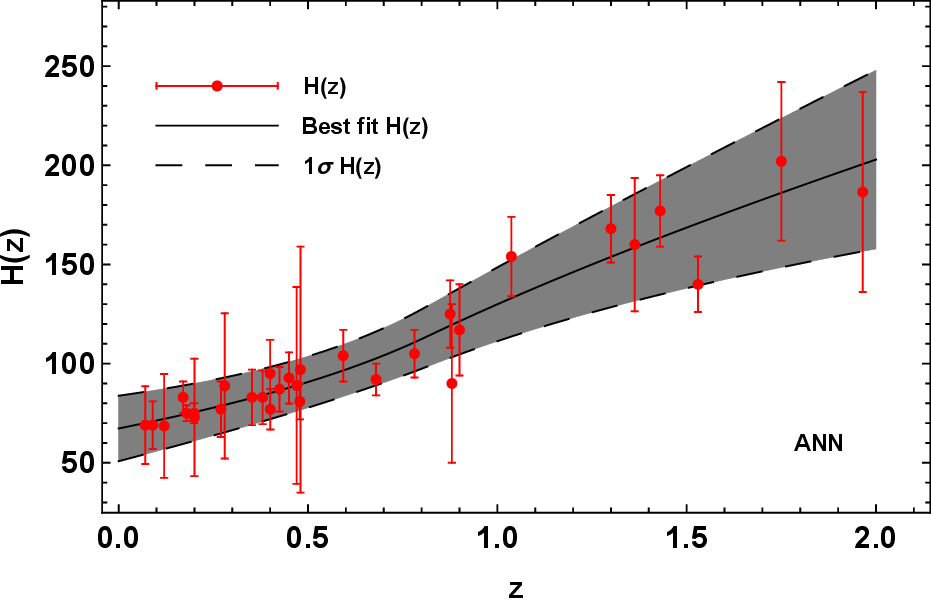}
\caption{\label{ANN} The distributions of the reconstructed functions $m_{\rm B}(z)$ (left) and $H(z)$ (right) with the corresponding $1\sigma$ errors with the ANN (black line), and the measurements of apparent magnitude from the Pantheon samples with Hubble parameter from the DA (red). }
\end{figure}

An ANN typically constitutes a deep learning algorithm that is structured into three principal layers: the input layer, the hidden layer, and the output layer. The input layer consists of $n$ nodes, where each node represents an independent variable, succeeded by $m$ interconnected hidden layers, culminating in the output layer equipped with activation functions as per the foundational design~\cite{Schmidhuber2015}. The ANN evaluates the error gradient derived from the training dataset, subsequently refining the weights of model and bias estimates during the backpropagation phase, advancing toward an optimal solution through the Adam optimization algorithm~\cite{Kingma2014}. The operational mechanism of ANN can be described in vectorized form, with further particulars delineated in Refs.~\cite{Wang2020,Clevert2015,LeCun2012}. We employ the publicly accessible code, titled Reconstructing Functions Using Artificial Neural Networks (ReFANN)\footnote{https://github.com/Guo-Jian-Wang/refann}~\cite{Wang2020}, to reestablish the relationship between the apparent magnitude $m_{{\rm B}}$ and the redshift $z$, as presented in the left panel of Fig.~\ref{ANN}. It is discernible that the uncertainties derived from the ANN-reconstructed function closely approximate those from the observational data. Furthermore, the reconstructed $1\sigma$ confidence level (CL) of $m_{{\rm B}}$ can be regarded as representative of the average observational error margin. The LD corresponding to the ADD derived from BAO data can be deduced through the smoothed function $m_{{\rm B}}(z)$ that has been reconstructed with an ANN. We refer to this hybrid method that combines the Binning method and ANN as method II. The BAO data points matched with the SNIa measurements at the identical redshift using the binning method or ANN in method II are presented in Tab.~\ref{tab:bao_measures}. Meanwhile, due to the limited amount of data on the Hubble parameter, it does not meet the redshift selection criteria, while it is matched with the BAO measurements. We utilize the ANN to obtain the continuous function of $H(z)$ by reconstructing 31 Hubble parameters, and the value of $H(z)$ at the redshift $z$ of the BAO are utilized to determine the ADD for the BAO observation. The reconstructed continuous function of $H(z)$ is represented in the right panel of Fig.~\ref{ANN}. It is evident that the compressed form of observed data, as well as the hybrid technique that combines the binning method with the ANN, is independent of any cosmological model.

\subsection{Methodology}
We employ the $\eta(z)$ function to examine potential violation from the CDDR at various redshifts through comparing the LD derived from SNIa observations with the ADD inferred from BAO measurements. The $\eta(z)$ function can be calculated using the subsequent formula:
\begin{equation}\label{eta}
\eta(z)={D_{\rm L}\over D_{\rm A}}{(1+z)^{-2}}\,.
\end{equation}
At any redshift, a value of $\eta(z)\neq 1$ signifies a discrepancy between CDDR and the observational data from astronomy. We utilize three distinct parameterizations for the function $\eta(z)$: the linear form P1: $\eta(z)=1+\eta_0z$, and two nonlinear forms P2: $\eta(z)=1+\eta_0z/(1+z)$, and P3: $\eta(z)=1+\eta_0\ln(1+z)$, taking into account the benefits of the $\eta(z)$ formulation, including its manageable one-dimensional phase space and its heightened responsiveness to observational data~\cite{Holanda2011}. The observed $\eta_{\rm obs}(z)$) is derived from Eq.~\ref{eta}.

The peak apparent magnitude $m_{\rm B}$, the measurements of  ${D_{\rm V}(z)/r_{\rm d}}$, $D_{\rm M}(z)/r_{\rm d}$, Hubble parameter $H(z)$, and redshift $z$, can be obtained directly from the observations of SNIa, BAO , and the passive evolutionary galaxies, independent of any cosmological model. From equations~\ref{muz}, \ref{DA1}, and \ref{DA2}, to obtain the value of $D_{\rm L}$ and $D_{\rm A}$, one needs to know the value of $M_{\rm B}$ and $r_{\rm d}$. Recent research has focused on deriving the values of $M_{\rm B}$ and $r_{\rm d}$ from a cosmological perspective, and the results indicate that different astronomical observations yield varied constraints on $M_{\rm B}$ and $r_{\rm d}$ (as discussed in the introduction of this work)~\cite{Dinda2023,Camarena2020b}. Consequently, when testing the CDDR, the use of specific prior values $M_{\rm B}$ and $r_{\rm d}$ to determine $D_{\rm L}$ and $D_{\rm A}$ introduces biases due to the adopted cosmological model and these preset values, rendering the method not entirely independent of the cosmological model. Therefore, it is imperative to employ new methods for testing the CDDR that are independent of  $M_{\rm B}$ and $r_{\rm d}$.

Given that the uncertainty to individual SNIa or BAO measurements is unrelated to $M_{\rm B}$ or $r_{\rm d}$, it is feasible to eliminate these parameters from the formula by performing an analytical marginalization over them during the analytical process. Following the procedure detailed in the study by~\cite{Xu2022}, we consider the fiducial values of $M_{\rm B}$ and $r_{\rm d}$ nuisance parameters for the determination of the DL and the ADD, subsequently marginalizing their impacts in the statistical analysis using a flat prior distribution. Consequently, the likelihood function represented by $\chi^{2}$ for the $D_{\rm V}(z)/r_{\rm d}$ data can be written as:
\begin{equation}
\label{chi2}
\chi^{2}(\eta_0, \kappa)= \sum_i^{N}\frac{{{\alpha_i^2 \over \beta_i^2}{\kappa}^2- 2 {\alpha_i \over \beta_i}{\kappa}+1  }}{\sigma^{2}_{{\eta_{{\rm obs},i}}}}\,,
\end{equation}
$N$ signifies the number of $D_{\rm V}(z)/r_{\rm d}$ data points from BAO utilized in this study. Here, $\alpha_i=\eta(z_i)$, $\beta_i=10^{({m_{{\rm B},i}\over 5}-5)}z_i^{1/2}c^{1/2}{(D_{\rm V}(z_i)/r_{\rm d})_{\rm obs}}^{-3/2}(1+z_i)^{-1}{H(z_i)}^{-{1/2}}$, $\kappa=10^{M_{\rm B} \over 5}\,r^{3/2}_{\rm d}$, and
\begin{equation}
\label{sigma01}
\sigma_{\eta_{\rm obs},i}^{2}=\left({\ln{10}\over {5}}{\sigma_{m_{{\rm B},i}}}\right)^2+\left({3\sigma_{{(D_{\rm V}(z_i)/r_{\rm d})_{\rm obs}}}\over2{{(D_{\rm V}(z_i)/r_{\rm d})_{\rm obs}}}}\right)^2+\left({\sigma_{H(z_i)}\over2{H(z_i)}}\right)^2.
\end{equation}

When employing $D_{\rm M}(z)/r_{\rm d}$ data, the likelihood function represented by $\chi^{\prime\,2}$ can be expressed as:
\begin{equation}
\label{chi2'}
\chi^{\prime\,2}(\eta_0, \kappa^{\prime})= \sum_j^{N^{\prime}}\frac{{{\alpha_j^{\prime\,2} \over \beta_j^{\prime\,2}}{\kappa}^{\prime\,2}- 2 {\alpha_j^{\prime} \over \beta_j^{\prime}}{\kappa^{\prime}}+1}}{\sigma^{\prime\,2}_{{\eta_{{\rm obs},j}^{\prime}}}}\,,
\end{equation}
$N^{\prime}$ denotes the count of $D_{\rm M}(z)/r_{\rm d}$ data points. Here, $\alpha_j^{\prime}=\eta(z_j)$, $\beta_j^{\prime}=10^{({m_{{\rm B},j}\over 5}-5)}{(D_{\rm M}(z_j)/r_{\rm d})_{\rm obs}}^{-1}(1+z_j)^{-1}$, $\kappa^{\prime}=10^{M_{\rm B} \over 5}\,r_{\rm d}$, and
\begin{equation}
\label{sigma02}
\sigma_{{\eta_{{\rm obs},j}^{\prime}}}^{\prime\,2}=\left({\ln{10}\over {5}}{\sigma_{m_{{\rm B},j}}}\right)^2+\left({\sigma_{{(D_{\rm M}(z_j)/r_{\rm d})_{\rm obs}}}\over{{(D_{\rm M}(z_j)/r_{\rm d})_{\rm obs}}}}\right)^2\,.
\end{equation}

\begin{figure}[htbp]
    \includegraphics[width=7.7cm]{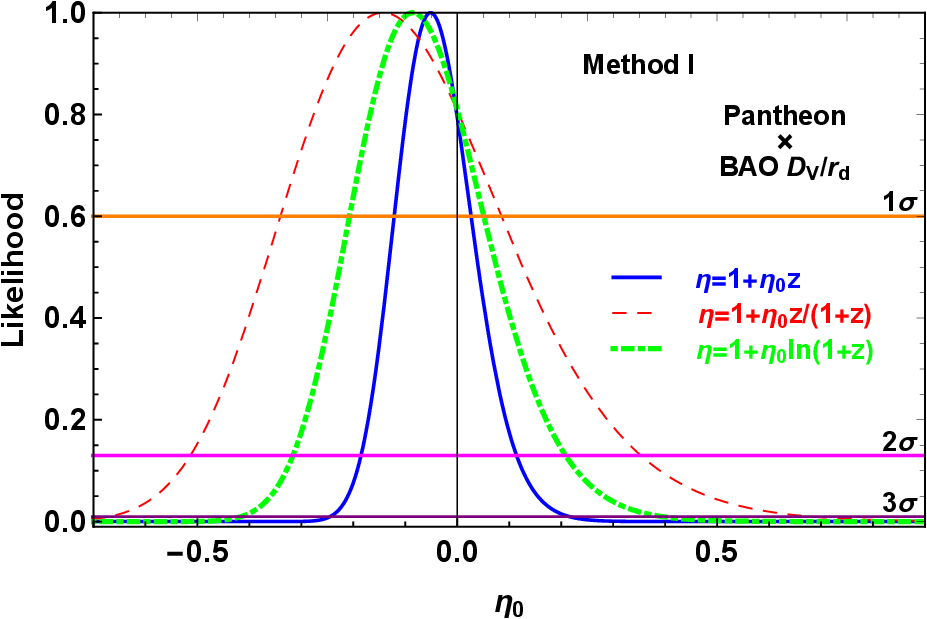}
	\includegraphics[width=7.7cm]{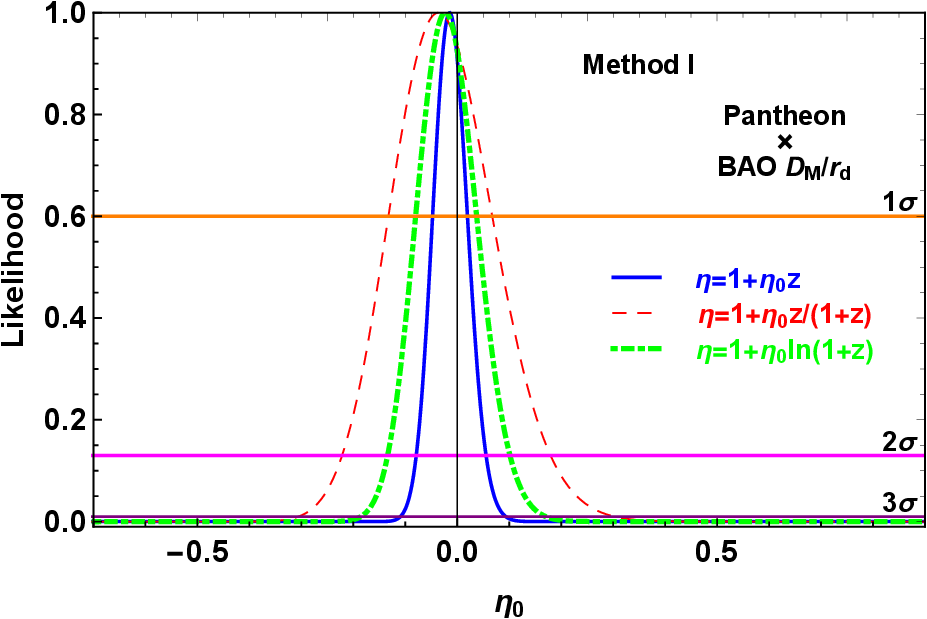}\\
	\hspace*{\fill}
	\includegraphics[width=7.7cm]{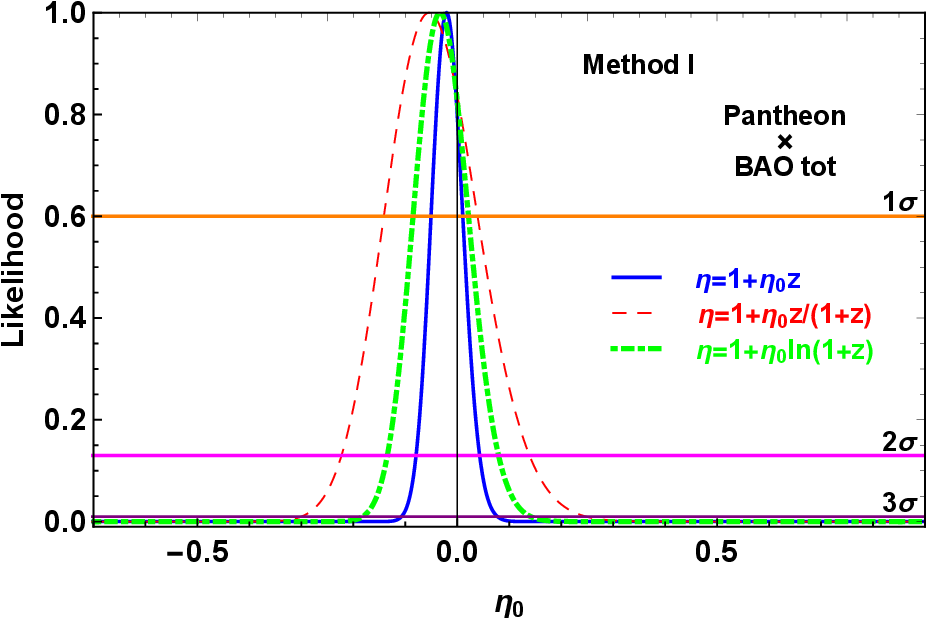}
	\hspace*{\fill}
	\caption{\label{figlikehood1} The likelihood distribution obtained from the $D_{\rm V}(z)/r_{\rm d}$ data (left), the $D_{\rm M}(z)/r_{\rm d}$ data (right), and the combined data (bottom) using method I with a flat prior.}
\end{figure}

\begin{figure}[htbp]
	
	\includegraphics[width=7.7cm]{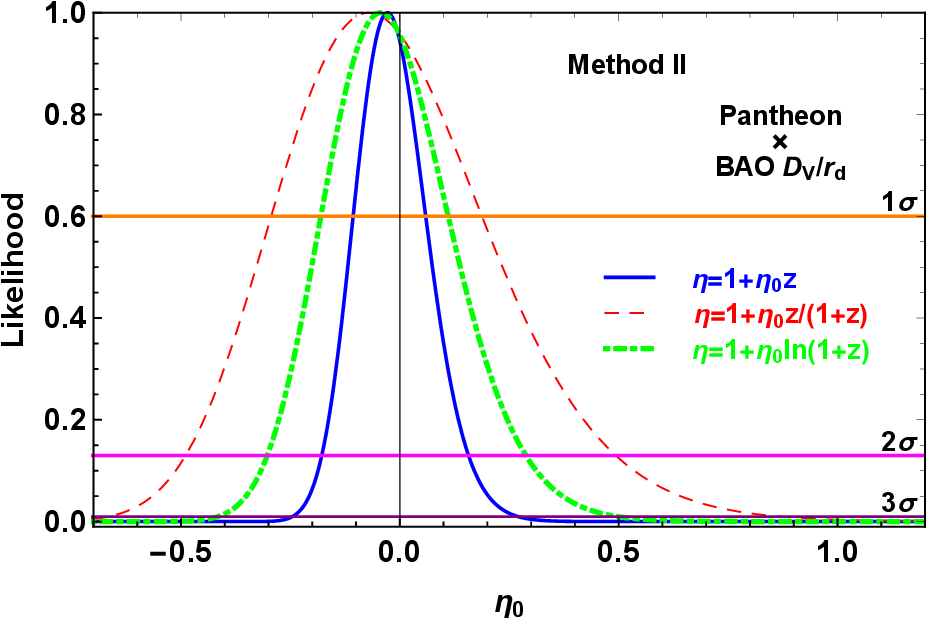}
    \includegraphics[width=7.7cm]{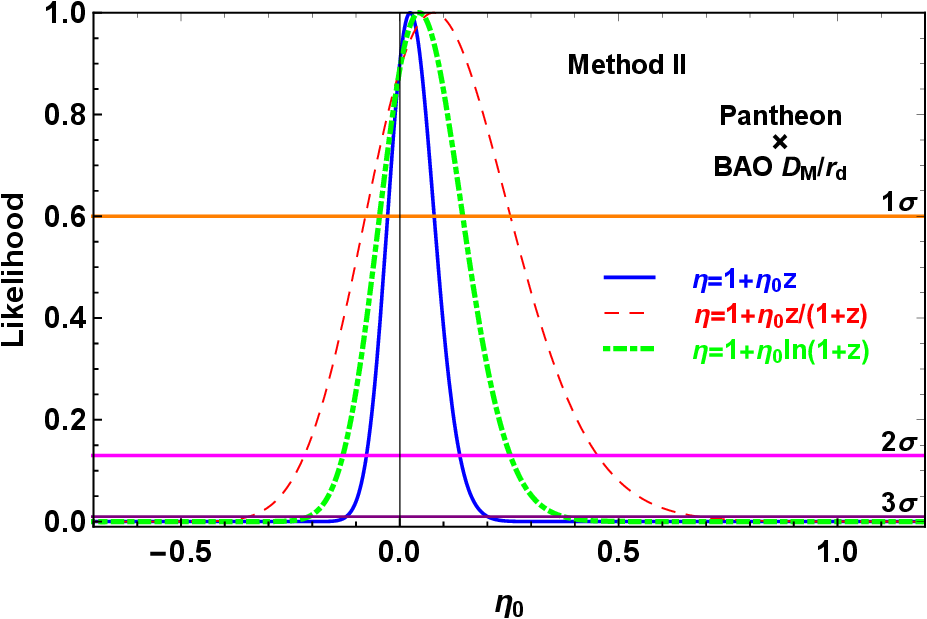}\\
	\hspace*{\fill}
	\includegraphics[width=7.7cm]{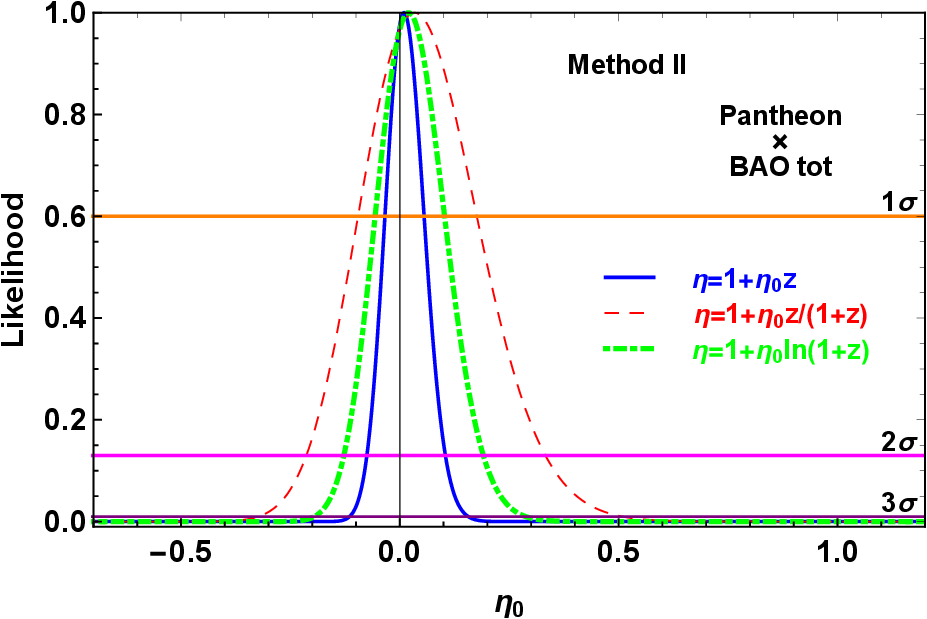}
	\hspace*{\fill}
	\caption{\label{figlikehood2} The likelihood distribution obtained from the $D_{\rm V}(z)/r_{\rm d}$ data (left), the $D_{\rm M}(z)/r_{\rm d}$ data (right), and the combined data (bottom) using method II with a flat prior.}
\end{figure}

In accordance with the methodology outlined in Refs.~\cite{Xu2022,Wang2023,Conley2010}, we marginalize analytically the likelihood
function over $\kappa$ and $\kappa^{\prime}$, assuming a flat prior on both $\kappa$ and $\kappa^{\prime}$. The marginalized $\chi^{2}$ in Eq.~\ref{chi2} and $\chi^{\prime\,2}$ in Eq.~\ref{chi2'} can be formulated as:
\begin{equation}
\label{chi3}
\chi_{\rm M}^{2}(\eta_0)= C-{B^2\over {A}}+\ln{A\over 2\pi}\,,
\end{equation}
\begin{equation}
\label{chi3'}
\chi_{\rm M}^{\prime\,2}(\eta_0)= C^{\prime}-{B^{\prime\,2}\over {A^{\prime}}}+\ln{A^{\prime}\over 2\pi}\,,
\end{equation}
where $A=\sum \alpha_i^2/(\beta_i^2{\sigma^{2}_{{\eta_{{\rm obs},i}}}})$, $B=\sum \alpha_i/(\beta_i{\sigma^{2}_{{\eta_{{\rm obs},i}}}})$, $C=\sum 1/{\sigma^{2}_{{\eta_{{\rm obs},i}}}}$, $A^{\prime}=\sum \alpha_j^{\prime\,2}/(\beta_j^{\prime\,2}{\sigma^{\prime\,2}_{{\eta_{{\rm obs},j}^{\prime}}}})$, $B^{\prime}=\sum \alpha_j^{\prime}/(\beta_j^{\prime}{\sigma^{\prime,2}_{{\eta_{{\rm obs},j}^{\prime}}}})$ and $C^{\prime}=\sum 1/{\sigma^{\prime\,2}_{{\eta_{{\rm obs},j}^{\prime}}}}$. Furthermore, we also contemplate the scenario where both datasets are jointly utilized for testing, with the corresponding chi-squared expressible as:
\begin{equation}
\label{chi4}
\chi_{\rm M,tot}^{2}(\eta_0)=\chi_{\rm M}^{2}(\eta_0)+\chi_{\rm M}^{\prime\,2}(\eta_0).
\end{equation}
The outcomes are depicted in Fig.~\ref{figlikehood1}, Fig.~\ref{figlikehood2} and Tab.~\ref{tablikelihood}.

It is clear that all parameters employed in the CDDR test are obtained directly from observational data, and the $\chi_{\rm M}^{2}$ in Eq.~\ref{chi3} and $\chi_{\rm M}^{\prime\,2}$ in Eq.~\ref{chi3'} are free from dependencies on parameters such as $M_{\rm B}$ and $r_{\rm d}$. As a result, through the process of analytical marginalization illustrated in Eq.\ref{chi2} and Eq.\ref{chi2'}, the variables $M_{\rm B}$ and $r_{\rm d}$ are removed from the fitting process. This enables a cosmological-model-independent parametric approach for testing the CDDR.
It is worth noting that, the CDDR
test in this work is also independent of the Hubble constant
$H_0$, as $r_{\rm d}\propto h^{-1}{\rm Mpc}$ is marginalized in the analysis. Thus, the parametric method used to test
CDDR is not only independent of cosmological model, but
also independent of the absolute magnitude $M_{\rm B}$ from SNIa
observation, sound horizon scale $r_{\rm d}$ of BAO measurements,
and Hubble constant $H_0$.

\begin{table}[htp]
	\begin{tabular}{|c|c|c|c|c|}
		\hline
		\scriptsize{parmetrization }  & \   P1: $1+\eta_0 {z}$\ \ &P2: $1+\eta_0 {z\over(1+z)}$  & \   P3: $1+\eta_0 {\ln(1+z)}$ \\
		\hline
		\scriptsize{$\eta_0^{\rm {A}\dag}$} &
		\scriptsize{${-0.051\pm^{0.078}_{0.070}\pm^{0.165}_{0.134}\pm^{0.266}_{0.193}}$}&
		\scriptsize{${-0.143\pm^{0.228}_{0.197}\pm^{0.496}_{0.370}\pm^{0.822}_{0.526}}$} &
		\scriptsize{${-0.086\pm^{0.137}_{0.121}\pm^{0.294}_{0.230}\pm^{0.479}_{0.330}}$} \\
		\hline
		\scriptsize{$\eta_0^{\rm {B}\dag}$} &
        \scriptsize{${-0.015\pm^{0.034}_{0.033}\pm^{0.070}_{0.064}\pm^{0.108}_{0.095}}$}&
		\scriptsize{${-0.036\pm^{0.104}_{0.096}\pm^{0.216}_{0.185}\pm^{0.339}_{0.269}}$} &
		\scriptsize{${-0.023\pm^{0.060}_{0.057}\pm^{0.124}_{0.111}\pm^{0.193}_{0.162}}$} \\
		\hline
		\scriptsize{$\eta_0^{\rm {C}\dag}$} &
        \scriptsize{${-0.021\pm^{0.031}_{0.030}\pm^{0.064}_{0.059}\pm^{0.098}_{0.087}}$}&
		\scriptsize{${-0.054\pm^{0.093}_{0.087}\pm^{0.193}_{0.168}\pm^{0.301}_{0.245}}$} &
		\scriptsize{${-0.033\pm^{0.055}_{0.052}\pm^{0.113}_{0.101}\pm^{0.174}_{0.149}}$} \\
		\hline
		\scriptsize{$\eta_0^{\rm {A}\ddag}$} &
        \scriptsize{${-0.028\pm^{0.087}_{0.079}\pm^{0.185}_{0.150}\pm^{0.298}_{0.217}}$}&
        \scriptsize{${-0.071\pm^{0.259}_{0.223}\pm^{0.564}_{0.418}\pm^{0.937}_{0.594}}$} &
        \scriptsize{${-0.044\pm^{0.154}_{0.137}\pm^{0.332}_{0.259}\pm^{0.541}_{0.371}}$} \\
		\hline
		\scriptsize{$\eta_0^{\rm {B}\ddag}$} &
        \scriptsize{${0.024\pm^{0.055}_{0.051}\pm^{0.113}_{0.100}\pm^{0.177}_{0.146}}$}&
        \scriptsize{${0.077\pm^{0.176}_{0.156}\pm^{0.374}_{0.295}\pm^{0.603}_{0.421}}$} &
        \scriptsize{${0.045\pm^{0.099}_{0.091}\pm^{0.208}_{0.175}\pm^{0.328}_{0.252}}$} \\
		\hline
		\scriptsize{$\eta_0^{\rm {C}\ddag}$} &
        \scriptsize{${0.010\pm^{0.046}_{0.044}\pm^{0.095}_{0.085}\pm^{0.147}_{0.124}}$}&
        \scriptsize{${0.034\pm^{0.142}_{0.129}\pm^{0.299}_{0.247}\pm^{0.474}_{0.357}}$} &
        \scriptsize{${0.020\pm^{0.082}_{0.077}\pm^{0.171}_{0.148}\pm^{0.267}_{0.215}}$} \\
		\hline
	\end{tabular}
\caption{The maximum likelihood estimation results for the parameterizations with the method I and method II. The $\eta_0$ is represented by the best fit value $\eta_{0,{\rm best}} \pm 1\sigma \pm 2\sigma \pm 3\sigma$ for each dataset. The superscripts A, B, and C represent the cases obtained from the $D_{\rm V}(z)/r_{\rm d}$ data, the $D_{\rm M}(z)/r_{\rm d}$ data, and the combined data, respectively. The superscript $\dag$ and $\ddag$ denote the results obtained from the method I and method II with the flat marginalization, respectively. }
\label{tablikelihood}
\end{table}

\begin{table*}[htb]
\label{tab:results}
\begin{center}
\begin{tabular}{ccccc} \hline \hline
Dataset used & P1: $1+\eta_0 {z}$\ \ &P2: $1+\eta_0 {z\over(1+z)}$  & \   P3: $1+\eta_0 {\ln(1+z)}$ \\ \hline
${\rm {Union2.1}}+{\rm {BAO(Marg)}}$~\cite{Wu2015} & ${-0.174{\pm^{0.253}_{0.199}}}$ & ${-0.409{\pm^{0.529}_{0.381}}}$    \\ [1ex]
${\rm {SNIa }}+{\rm {BAO}(Marg)}$~\cite{Xu2020} & ${-0.07{\pm{0.12}}}$ & ${-0.20{\pm{0.27}}}$  & ${-0.12{\pm{0.18}}}$  \\ [1ex]
${\rm {SNIa }}+{\rm {BAO}(Marg)}$~\cite{Wang2024} & ${0.041{\pm^{0.123}_{0.109}}}$ & ${0.082{\pm^{0.246}_{0.214}}}$  & ${0.059{\pm^{0.174}_{0.159}}}$  \\ [1ex]
${\rm {SNIa }}+{\rm {BAO}(Marg)}$~\cite{Xu2022} & ${-0.037{\pm^{0.110}_{0.097}}}$ & ${-0.101{\pm^{0.269}_{0.225}}}$  & ${-0.061{\pm^{0.173}_{0.149}}}$  \\ [1ex]
${\rm {SNIa }}+{\rm {QSO}(Marg)}$~\cite{Yang2024} & ${-0.044{\pm^{0.049}_{0.046}}}$ & ${-0.256{\pm^{0.137}_{0.121}}}$  & ${-0.114{\pm^{0.084}_{0.076}}}$  \\ [1ex]
\hline \hline
\end{tabular}
\caption[]{Summary of the constraints on parameter $\eta_0$ with different data sets. ``Marg'' represents the results obtained by marginalizing certain parameters with a flat prior.}
\label{otherobser}
\end{center}
\end{table*}

\section{Results and Analysis}

In the case of result obtained with the method I and method II, the CDDR is consistent with the BAO observation concerning the five measurements of  $D_{\rm V}(z)/r_{\rm d}$, nine measurements of $D_{\rm M}(z)/r_{\rm d}$, and the combination of these datasets at $1\sigma$ CL for all three parameterizations P1, P2, and P3. The compressed form of the Pantheon SNIa sample used in Method I, which utilizes more actual SNIa data, generally yields tighter constraints compared to Method II, which employs hybrid method by combining the binning method with the artificial neural network (ANN). This underscores the advantage of Method I, as it enables observations to impose a more stringent constraint on the CDDR due to the incorporation of a greater number of SNIa data, which facilitates the derivation of more accurate values of $m_{\rm B}$. Furthermore, it can be inferred that among the three parameterizations, P1 imposes the most rigorous constraints on the parameter $\eta_0$ among the three parameterizations.

To explore the capability of the measurements of $D_{\rm V}(z)/r_{\rm d}$ of the BAO observations, it is necessary to compare our results with the previous constraints on $\eta_0$ from different data sets of SNIa and BAO. These BAO measurements improve the accuracy of $\eta_0$ about 60\% at the $1\sigma$ CL relative to the Union2.1+BAO observations, where the dimensionless Hubble constant $h$ was marginalized using a flat prior~\cite{Wu2015}. The bounds on $\eta_0$ are approximately 30\% stricter than those derived from the Pantheon compilation and the BOSS DR12 BAO data in the redshift region $0.31\leq{z}\leq0.72$~\cite{Xu2020}, as well as from 13 transverse BAO measurements obtained using the SDSS alongside the Pantheon SNIa samples, where the $M_{\rm B}$ and $r_{\rm d}$ were marginalized~\cite{Wang2024}. The constraints are approximately 20\% more restrictive than those obtained from the five $D_{\rm M}(z)/r_{\rm d}$ data points of BAO observations using the eBOSS DR16 quasar dataset combined with the Pantheon SNIa sample~\cite{Xu2022}, and are comparable to those derived from the Pantheon sample and compact radio quasars (QSO) measurements~\cite{Yang2024}, where the variables $M_{\rm B}$ and $r_{\rm d}$ is marginalized. Thus, the measurement $D_{\rm V}(z)/r_{\rm d}$ of the BAO observations provides an effective tool to test CDDR. It is worth noting that the parametric method used to test
CDDR in our analysis is independent not only of cosmological model, but
also of the absolute magnitude $M_{\rm B}$ from SNIa
observation, the sound horizon scale $r_{\rm d}$ of BAO measurements,
and the Hubble constant $H_0$. This independence ensures that our test results are robust and not biased by specific assumptions or prior values of these parameters.

For the case of the four BAO data points from DESI included, the renewed $D_{\rm M}(z)/r_{\rm d}$ measurements of BAO observations result in the precision enhancement of $\eta_0$ by approximately 50\% compared to the eBOSS DR16 quasar dataset~\cite{Xu2022}. When combining the measurements of  $D_{\rm V}(z)/r_{\rm d}$ with the measurements of  $D_{\rm M}(z)/r_{\rm d}$  of BAO observations, the constraints on $\eta_0$ become approximately 60\% stricter than those obtained from the measurements of  $D_{\rm V}(z)/r_{\rm d}$, and 70\% stricter than those obtained from eBOSS DR16 quasar dataset~\cite{Xu2022}. As a result, the inclusion of the recently released DESI BAO data, along with the combination of the measurements of $D_{\rm V}(z)/r_{\rm d}$ and $D_{\rm M}(z)/r_{\rm d}$  from BAO observations markedly enhances the constraints imposed on the CDDR. This improvement emphasizes the importance of ongoing efforts to expand and refine observational datasets.

\section{Conclusion}
The CDDR is crucial in astronomy and contemporary cosmology, with deviations potentially signaling new physical phenomena. SNIa and BAO measurements are effective tools for validating the CDDR. Since the measurements of the ratio of spherically-averaged distance  $D_{\rm V}(z)$ to sound horizon scale $r_{\rm d}$ encompasses comoving distance $D_{\rm M}(z)$ information, it provides the opportunity to extract ADD from the measurements of $D_{\rm V}(z)/r_{\rm d}$ for CDDR testing. The recent BAO data has been renewed by the DESI collaboration, incorporating two data points of $D_{\rm V}(z)/r_{\rm d}$ measurements and four data points of $D_{\rm M}(z)/r_{\rm d}$ measurements. In this study, we conduct a comprehensive test of the CDDR by integrating the latest measurements of BAO from the DESI collaboration with the Pantheon sample of SNIa. Our analysis aims to test the validity of CDDR in a manner independent of both cosmological models and the specific prior values of the absolute magnitude $M_{\rm B}$ of SNIa and the sound horizon scale $r_{\rm d}$ from BAO measurements.

To achieve this, we first utilize the measurements of $D_{\rm V}(z)/r_{\rm d}$ to obtain the ADD. By employing an ANN to reconstruct the Hubble parameter data acquired through the DA techniques, we derive ADD values from $D_{\rm V}(z)/r_{\rm d}$ BAO observations. A compressed version of the Pantheon sample and a hybrid method by combining the ANN and the binning method are employed to match the LD with the ADD data points at the identical redshift.
To avoid the potential biases arising from the specific prior values of the $M_{\rm B}$ of SNIa and the $r_{\rm d}$ from BAO measurements in CDDR test, we introduce a new variable $\kappa\equiv10^{M_{\rm B} \over 5}\, r_{\rm d}^{3 \over 2} $ and treat it as a nuisance parameter, marginalizing its influence in our statistical analysis using a flat prior distribution. Our results show that the CDDR is consistent with the astronomic observations and the measurement of $D_{\rm V}(z)/r_{\rm d}$  from BAO observations provides an effective tool to test CDDR. The compressed Pantheon SNIa sample used in Method I, incorporating more actual data, typically provides tighter constraints than Method II, which combines binning with an artificial neural network (ANN). The parametric method used to test
CDDR in our analysis is independent not only of cosmological model, but
also of the absolute magnitude $M_{\rm B}$ from SNIa
observation, the sound horizon scale $r_{\rm d}$ of BAO measurements,
and the Hubble constant $H_0$.

Furthermore, we update the $D_{\rm M}(z)/r_{\rm d}$ dataset from BAO observations to test the CDDR using the latest BAO findings from the DESI collaboration. By combining the $D_{\rm V}(z)/r_{\rm d}$ and $D_{\rm M}(z)/r_{\rm d}$ data sets, we obtain more precise constraints on the CDDR. We show that the inclusion of the recently released DESI BAO data, along with the combination of $D_{\rm V}(z)/r_{\rm d}$ and $D_{\rm M}(z)/r_{\rm d}$ measurements from BAO observations markedly improves the constraints on the CDDR, affirming that BAO observations can serve as a powerful tool for testing the CDDR through cosmological-model-independent method. Therefore, updating and merging different types of BAO observational data is crucial for effectively enhancing the constraints on the CDDR.

\begin{acknowledgments}
We very much appreciate the helpful comments and
suggestions from anonymous referees, and we would also
like to thank Puxun Wu for helpful discussions. This work was supported by the National Natural Science Foundation of China under Grants No. 12375045, No. 12305056, No. 12105097 and No. 12205093, the
Hunan Provincial Natural Science Foundation of China
under Grants No. 12JJA001 and No. 2020JJ4284, the Natural Science Research Project of Education
Department of Anhui Province No. 2022AH051634, and the
Science Research Fund of Hunan Provincial Education
Department No. 21A0297.

\end{acknowledgments}

\end{document}